# A wide-orbit giant planet in the high-mass b Centauri binary system

Markus Janson*[1], Raffaele Gratton[2], Laetitia Rodet[3], Arthur Vigan[4], Mickaël Bonnefoy[5], Philippe Delorme[5], Eric E. Mamajek[6], Sabine Reffert[7], Lukas Stock[7], Gabriel-Dominique Marleau[8,9,10], Maud Langlois[11], Gaël Chauvin[12,5], Silvano Desidera[2], Simon Ringqvist[1], Lucio Mayer[13], Gayathri Viswanath[1], Vito Squicciarini[2,14], Michael R. Meyer[15], Matthias Samland[1], Simon Petrus[5], Ravit Helled[13], Matthew A. Kenworthy[16], Sascha P. Quanz[17], Beth Biller[18], Thomas Henning[10], Dino Mesa[2], Natalia Engler[17], Joseph C. Carson[19]

[1]Department of Astronomy, Stockholm University, 10691, Stockholm, Sweden

[2]INAF Osservatorio Astronomico di Padova, vicolo dell'Osservatorio 5, 35122, Padova, Italy

[3]Cornell Center for Astrophysics and Planetary Science, Department of Astronomy, Cornell University, Ithaca, NY 14853, USA

[4]Aix-Marseille Université, CNRS, CNES, LAM (Laboratoire d'Astrophysique de Marseille) UMR 7326, 13388, Marseille, France

[5]Univ. Grenoble Alpes, CNRS, IPAG, F-38000 Grenoble, France

[6]Jet Propulsion Laboratory, California Institute of Technology, 4800 Oak Grove drive, Pasadena CA 91109, USA

[7]Landessternwarte, Zentrum für Astronomie der Universität Heidelberg, Königstuhl 12, 69117 Heidelberg, Germany

[8]Institut für Astronomie und Astrophysik, Universität Tübingen, Auf der Morgenstelle 10, D-72076 Tübingen, Germany

[9]Physikalisches Institut, Universität Bern, Gesellschaftsstr. 6, CH-3012 Bern, Switzerland

[10]Max-Planck-Institut für Astronomie, Königstuhl 17, D-69117 Heidelberg, Germany

[11]CRAL, UMR 5574, CNRS, Université Lyon 1, 9 avenue Charles André, 69561 Saint Genis Laval Cedex, France

[12]Unidad Mixta Internacional Franco-Chilena de Astronomía, CNRS/INSU UMI 3386 and Departamento de Astronomía, Universidad de Chile, Casilla 36-D, Santiago, Chile

[13]Center for Theoretical Physics and Cosmology, Institute for Computational Science, University of Zurich, Winterthurerstrasse 190, 8056, Zurich, CH

[14]Department of Physics and Astronomy "Galileo Galilei", vicolo dell'Osservatorio 3, 35122, University of Padova, Italy

[15]Department of Astronomy, University of Michigan, 1085 S. University Ave, Ann Arbor MI 48109, USA

[16]Leiden Observatory, Leiden University, Postbus 9513, 2300 RA Leiden, The Netherlands

[17]ETH Zurich, Institute for Particle Physics and Astrophysics, Wolfgang-Pauli-Strasse 27, 8093 Zurich, Switzerland

[18]SUPA, Institute for Astronomy, Royal Observatory, University of Edinburgh, Blackford Hill, Edinburgh EH93HJ, UK

[19]College of Charleston, Department of Physics & Astronomy, 66 George St, Charleston, SC 29424 USA

**Planet formation occurs around a wide range of stellar masses and stellar system architectures[1]. An improved understanding of the formation process can be achieved by studying it across the full parameter space, particularly toward the extremes. Earlier studies of planets in close-in orbits around high-mass stars have revealed an increase in giant planet frequency with increasing stellar mass[2] until a turnover point at 1.9 solar masses, above which the frequency rapidly decreases[3]. This could potentially imply that planet formation is impeded around more massive stars, and that giant planets around stars exceeding 3 solar masses may be rare or non-existent. However, the methods used to detect planets in small orbits are insensitive to planets in wide orbits. Here we demonstrate the existence of a planet at 560 times the Sun-Earth distance from the 6—10 solar mass binary b Centauri through direct imaging. The planet-to-star mass ratio of 0.10—0.17% is similar to the Jupiter-Sun ratio, but the separation of the detected planet is ~100 times wider than that of Jupiter. Our results show that planets can reside in much more massive stellar systems than what would be expected from extrapolation of previous results. The planet is unlikely to have formed in-situ through the conventional core accretion mechanism[4], but might have formed elsewhere and arrived to its present location through dynamical interactions, or might have formed via gravitational instability.**

The observations of the b Centauri (b Cen, HR 5471, HIP 71865; b Cen ≠ β Cen) system were acquired as part of the B-star Exoplanet Abundance Study (BEAST)[5], which surveys massive stars in the Scorpius-Centaurus (Sco-Cen) young stellar association with high-contrast imaging for direct detection of planetary companions. We acquired an original epoch observation in 2019 (Fig. 1), in which we identified three faint point sources around b Cen, where the brightest one had interesting near-infrared colours similar to previously imaged planetary companions. Generally, a faint point source can either be a planet in orbit around the target star, or a chance alignment of a background star. The two scenarios can be distinguished by assessing whether the point source shares a common proper motion with the target star, in which case it can be established as being physically bound to the target. We therefore scheduled a follow-up observation of b Cen, which was executed in 2021 (Extended Data Fig. 1). In addition, we found that the candidate planet appeared as a point source in archival observations from a direct imaging campaign taken in 2000[6]. The candidate had been noted in the survey report[6], but all candidates fainter than 13 mag in the J-band were assumed to be background contaminants in that report, so it was discarded without further follow-up. All of the data we have collected confirm at >7.3σ significance that the candidate shares a common proper motion with b Cen (Extended Data Fig. 2), and furthermore, there is clear evidence for orbital motion consistent with the expected orbital speed around the central stellar mass (Extended Data Figs 2, 3). The colours of the companion are also consistent with young objects of planetary masses (Extended Data Fig. 4). This collected body of evidence firmly establishes that the candidate is a directly imaged exoplanet, physically bound to the b Cen system. The two other faint point sources seen in the BEAST images are confirmed as background stars (Extended Data Fig. 2).

The b Cen system consists of a close pair of stars. The more massive star is named b Cen A and has a spectral type of B2.5V[7], corresponding to an effective temperature of approximately 18,000 K. The second star b Cen B has been seen through its dynamical influence on its primary star[8], but there is no full orbital characterization of the system, so its properties are uncertain. Because of its circumbinary nature, we will refer to the detected planet as "b Cen (AB)b", where (AB) denotes that it orbits both of the stellar A and B components. We estimate the mass of the b Cen AB stars with isochronal models[9], using the system age of 15 ± 2 Myr based on a 99.8% probability membership of the Upper Centaurus Lupus association, and its parallax-based distance of 99.7 ± 3.1 parsecs[10]. In this way, we find a total mass of 6—10 solar masses for the central pair (Table 1). This is 2—4 times higher than for the host stars any other confirmed planets: HD 106906 AB, which hosts a directly imaged planet, is the next highest-mass binary host with a total mass of 2.7 solar masses[11], and in terms of single stars, the highest mass verified planet host stars in radial velocity surveys have masses up to 3 solar masses[3].

Based on isochronal fits to the photometric data, b Cen (AB)b has a luminosity of $1.0 \times 10^{-4}$ solar luminosities, consistent with theoretical expectations for a 15 Myr super-jovian planet. The luminosity can be used to derive an estimate for the planetary mass, although there is a degeneracy with the entropy at the end of formation, which is associated with a considerable uncertainty[12]. However, for an object in the age and luminosity range of b Cen (AB)b, the impact is modest. We perform a Markov chain Monte Carlo test with the BEX-Cond exoplanet cooling models[13] with a wide range of initial entropies (ranging from "cold-start" to "hot-start") to derive a mass of 10.9 ± 1.6 $M_{Jup}$ (Extended Data Fig. 5). Warm- and hot-start conditions are favoured by the model fitting. The mass estimations are very similar to those of the imaged exoplanet HIP 65426 b when subjected to the same analysis[13], which is as expected since the two planets have both ages and luminosities within 10% of each other. With a planetary mass of approximately 11 $M_{Jup}$, the mass ratio of b Cen (AB)b to the central binary is only 0.10—0.17% (Fig. 2). This is similar to the mass ratio between Jupiter and the Sun. The mass ratio of a planet to the host star system it orbits is thought to provide clues to its formation. For example, there is a well known bimodal distribution seen among companions to Sun-like stars, where one population increases in frequency downward from a ~1% mass ratio, while another population increases upward of ~10%, with a large unpopulated range in between[14]. The divide represents differences between the planet formation and stellar companion formation scenarios. In this context, b Cen (AB)b would fall firmly in the planet-formation regime.

Thanks to the recovered epoch from 2000, our observational baseline spans more than two decades. In combination with the relatively fast orbital motion (despite the large semi-major axis) facilitated by the high central mass, this means that we can measure statistically significant orbital motion, which in turn means that we can put initial constraints on the orbital properties. We have run an orbit-fitting code[15] suitable for orbits with limited coverage of the full orbital period (see 'Methods' section). In this way, we find that the inclination $i$ lies in the range of 128°—157° within a 68% confidence interval, meaning it is intermediate between an edge-on (90°) and a face-on (0° or 180°) orbital orientation, possibly closer to face-on; and the eccentricity $e$ is low, $e$ < 0.40 at 68% confidence.

Both direct imaging studies[16,17] and indirect radial velocity studies[2,3] have shown a clear trend of massive planetary companions becoming increasingly abundant around increasingly massive stars, probably as a result of an increasing circumstellar disc mass for more massive stars[4]. However, in radial velocity surveys, the frequency of planet occurrence turns over at around 1.9 solar masses, going to effectively zero above 3 solar masses[3]. Since most previous large-scale exoplanet surveys have generally centred their attention around more Sun-like stars, the >3 solar mass range has not been systematically probed with direct imaging prior to BEAST, and only recently explored in radial velocity[3]. This turnover might be related to the increased levels of high-energy radiation emitted from massive stars, causing faster disc evaporation and therefore a shorter disc lifetime[18]. If the disc typically evaporates faster than giant planets can form in it, this would naturally explain a decreasing planetary frequency past a certain stellar mass[19]. Alternatively, the inside-out evolution of disc dissipation might allow planets on wide orbits to form around massive stars, but prevent them from migrating to smaller separations[20]. Since radial velocity has a strong detection bias toward close-in planets, it would be blind to such migration-halted planets. Direct imaging has the opposite bias, with primary sensitivity to wider planets, and could therefore distinguish between the formation-halted and migration-halted scenarios outlined above. The planet b Cen (AB)b is a potential representative of a migration-halted planet population. Since protoplanetary disks around young massive stars can reach sizes in the range of thousands of au[21] (1 au is equal to the Sun-Earth semi-major axis), the b Cen AB(b) separation of 550 au is possible in this context.

The bulk of the known giant planet population is consistent with having been formed through core accretion, in which solids in a young circumstellar disc accumulate into cores onto which gas rapidly accretes once the core reaches a critical mass[22]. However, critical core build-up is expected to be exceedingly difficult at separations much wider than the circumstellar snow line[4], although core formation through so-called pebble accretion may facilitate formation at wider separations than the classical core formation scenario of planetesimal collisions[23]. This issue may be particularly important for massive host stars, where the gas disc dissipates faster than for Sun-like stars, leaving less time for the core to reach the critical mass for gas accretion. It is therefore unlikely that b Cen (AB)b formed in-situ by core accretion. One option could be that the planet formed closer in toward the parent stars, and was subsequently ejected to a larger orbit through dynamical interactions with other planets in the system of similar mass or higher. However, in this case, we would expect a high eccentricity for the planetary orbit, as well as additional companions in the system to have caused the scattering. By contrast, the measured eccentricity is modest or low, and no companions of similar mass or higher than b Cen (AB)b are visible in the images. Consequently, such companions of equal mass to b Cen (AB)b can be excluded by 5σ confidence down to a separation of 25 au, i.e., >20 times smaller than the b Cen (AB)b orbit. Another source of scattering might be the central binary, either through close encounters or through mean motion resonances[24,25]. If the current binary configuration is primordial, such interactions would be very unlikely, since the planet separation is >100 times larger than the binary separation. However, the stellar separation might have been larger at the time of formation, and subsequently migrated inwards. There are so far no clear differences between the wide planetary populations of single and binary stars in statistical studies[26]. However, future studies including b Cen (AB)b and other new detections could potentially reveal such differences, which would imply that binary interactions might have played a role in the formation of b Cen (AB)b.

Alternatively, a giant planet could form directly from the circumstellar gas disc through gravitational instability[27]. This might be a particularly important mechanism in the context of massive stars, as the full process of formation can in principle occur in a few orbital timescales, i.e. $\sim 10^4$ years in the case of b Cen (AB)b. Since this is much faster than the $\sim 10^6$ years required for core accretion, the instability mechanism is less sensitive to the rapid dispersion timescales of discs around massive stars. Meanwhile, the migration timescale is independent of the formation mechanism. Hence, a scenario in which b Cen (AB)b formed rapidly through gravitational instability close to its present orbit but was prevented from migrating substantially inward due to rapid dispersion is a possible explanation for why it is observed in its current environment. During the window between formation and disc dispersal, the net migration might even have been in the outward direction, as is sometimes seen in simulations of disc fragmentation[28]. The relatively high initial entropy of the planet implied by our analysis might speak in favour of gravitational instability as a formation scenario, although other mechanisms cannot be excluded on this basis[29]. Theoretical models predict that discs around more massive stars are more likely to fragment as a result of more vigorous mass accretion[30], which further supports the interpretation of b Cen (AB)b as disc instability planet. As yet another option, the planet could have formed as part of a separate stellar system and subsequently ejected, or it could have formed in isolation. At a later stage, it would then have been gravitationally captured by the b Cen system. Planet transfers between stellar hosts are possible in young star-forming regions where the stellar density is high[31,32]. However, since high stellar density environments tend to end up as open clusters that remain clustered for hundreds of Myr, the fact that b Cen currently resides in a non-bound kinematical association at an age of only 15 Myr implies that it was never part of any very high density environment. In addition, captured planets should have high eccentricities in general, $f(e) = 2e$ where $e$ is the eccentricity and $f$ is the distribution of eccentricities in a captured population[33]. The probability of acquiring $e < 0.40$ is approximately 17% in this context, so while the gravitational capture scenario cannot be excluded, it is mildly disfavoured by the existing data. A wide and relatively low-eccentricity orbit is not unique to b Cen (AB)b, but is also seen in other directly imaged planets; most notably, the four HR 8799 planets[34]. However, at semi-major axes ranging ~16—70 au, the HR 8799 planets still reside at an order of magnitude smaller orbital separations than b Cen (AB)b at 556 au. The closest known analogue to b Cen (AB)b may be HD 106906 (AB)b[35], which is another very wide (~650 au) circumbinary planet, although in a substantially less massive system containing two roughly equal-mass stars of 1.31-1.37 solar masses, and a system mass of 2.7 solar masses[11]. We show b Cen (AB)b in the context of the wider exoplanetary population in Fig. 2. So far, there are no other known systems quite like it.

Our measurements of the orbital properties of b Cen (AB)b disfavour a dynamically violent past, favouring instead a formation close to its present location with little subsequent orbital evolution. Since core accretion is challenging at such large separations, disc instability might represent a more probable formation scenario. Regardless of the specific formation mechanism, the discovery of b Cen (AB)b shows that the statistical ~3 solar mass upper stellar limit for hosting giant exoplanets within ~5 au implied from radial velocity measurements[3] cannot be extrapolated to the full system architecture. Stars and stellar systems up to at least 6—10 solar masses can host giant planets on wide orbits.

**Acknowledgements** We thank M. Ireland for input regarding SUSI interferometry. The results presented here are based on observations collected at the European Organisation for Astronomical Research in the Southern Hemisphere under ESO programme 1101.C-0258. The work has made use of the SPHERE Data Centre, jointly operated by OSUG/IPAG (Grenoble), PYTHEAS/LAM/CeSAM (Marseille), OCA/Lagrange (Nice), Observatoire de Paris/LESIA (Paris), and Observatoire de Lyon/CRAL, and supported by a grant from Labex OSUG@2020 (Investissements d'avenir – ANR10 LABX56). The study made use of CDS and NASA-ADS services. M.J. acknowledges support from the Knut and Alice Wallenberg Foundation (KAW). G.-D.M. acknowledges the support of the DFG priority program SPP 1992 "Exploring the Diversity of Extrasolar Planets" (MA 9185/1-1) and from the Swiss National Science Foundation under grant BSSGI0_155816 "PlanetsInTime". Parts of this work have been carried out within the framework of the NCCR PlanetS supported by the Swiss National Science Foundation. A.V. acknowledges funding from the European Research Council (ERC) under the European Union's Horizon 2020 research and innovation programme (grant agreement No. 757561). Part of this research was carried out at the Jet Propulsion Laboratory,

California Institute of Technology, under a contract with the National Aeronautics and Space Administration (80NM0018D0004).

**Author contributions** M.J. is PI for the BEAST survey and led the management, observation preparations, analysis and manuscript writing. R.G., A.V., M.B., G.-D.M., S.Ri., G.V., V.S. and S.P contributed to the data analysis and plots. L.R. led the orbital fitting. P.D. led the data reduction. R.G., E.E.M., S.Re., L.S., G.-D.M. and S.D. contributed to the stellar and planetary characterization. M.L. contributed to the observation preparations. G.C. assisted with the project management. L.M., M.R.M. and R.H. contributed to the formation discussion. All co-authors assisted with the manuscript writing.

**Author information** Reprints and permissions information is available at www.nature.com/reprints. The authors declare no competing financial interests. Correspondence and requests for materials should be addressed to M.J. (markus.janson@astro.su.se).

**METHODS**

**Observations and data reduction.** The target b Cen was observed with SPHERE[36] at the Very Large Telescope located in Paranal, Chile, on 20 Mar 2019 and on 10 Apr 2021 as part of the BEAST[5] survey. The 2019 observations were executed in the so-called IRDIFS-EXT mode[37]. In the IRDIFS-EXT mode, light is split up spectrally with a dichroic, such that light in the YJH-band range is recorded by the IFS arm of SPHERE, and light in the K-band range is recorded by the IRDIS arm. Since the IFS field of view is only 1.7×1.7 arcseconds, it does not contain the planetary companion (located at 5.4 arcseconds separation) and is therefore not used in this analysis. IRDIS records simultaneous images across a 12×12 arcsecond field of view in two separate wavelength bands[38], which in the 2019 epoch were the K1 (2110 nm) and K2 (2251 nm) bands. The 2021 observations were acquired with IRDIS in a stand-alone mode, using the J2 (1190 nm) and J3 (1273 nm) bands. The standard SPHERE coronagraph called "N-ALC-YJH-S" with an inner working angle of approximately 0.1 arcseconds was used during most of the observing sequence, except for a few frames at the beginning and end of the sequence that were taken with a neutral density filter in the beam, for unsaturated photometric referencing of the unresolved central stellar pair.

Data reduction was performed using the SPHERE Data Center[39] software, using the SpeCal[40] high-contrast algorithm package. The standard BEAST reduction procedure[5] was performed, including classical angular differential imaging (cADI)[41], template locally optimized combination of images (TLOCI)[42], and a pure image rotation and combination algorithm. Due to its large separation (~5.35 arcseconds) from the central stars, the planet b Cen (AB)b was detected in all reductions, and in all bands during both epochs. Two faint background stars were also identified in the more sophisticated of the reduction schemes. Negative injection in the TLOCI scheme was used to derive photometry and astrometry of the planet and the background stars for robust estimations of the uncertainties involved. Calibration of astrometric parameters such as the pixel scale and the true North alignment was performed using cluster observations following the standard Data Center calibration scheme[43]. The astrometric properties of all the points sources in the field are shown in Extended Data Table 1, and the photometric properties are shown in Extended Data Table 2.

**Stellar system analysis.** The distance to the b Cen system is 99.7 ± 3.1 parsecs, based on the parallax from the latest data release from the *Gaia* satellite[10]. This value is well consistent with others from previous releases and from *Hipparcos*[44]. However, continued evaluation of the distance in the future is relevant, since it is not yet clear to which extent the exact parallax could be affected by the stellar binary motion, which could in principle have a similar orbital period as the parallactic period of one year. The proper motion of the system is insensitive to such short-period events, and is measured as 29.83 ± 0.37 mas/yr Westward and 31.91 ± 0.52 mas/yr Southward[10]. We base the age estimate of the system on the membership of b Cen in the Upper Centaurus Lupus (UCL) sub-group of the Scorpius-Centaurus region[45], and its specific location within this region. According to the BANYAN-Σ[46] tool, the membership probability of b Cen to UCL is 99.8%. While the UCL region exhibits a non-negligible age spread, particularly toward its edges, b Cen is located in a large uniform age area in an age map[47] of the region, with a mean age of 15 Myr. Indeed, based on the standard deviation in a circular area with a 10° diameter centered on b Cen in the map, the age scatter in that part of UCL is only ±1 Myr. Hence, we conclude that the b Cen age uncertainty is dominated by the intrinsic UCL mean age uncertainty[47], leading to an age estimate of 15 ± 2 Myr.

For estimating the effective temperature of the primary star, we use a combination of literature values[48,49] to acquire $T_{eff}$ = 18310 ± 320 K. Several extinction estimates based on existing literature[48,49,50] give $E(B-V)$ color excesses in the vicinity of 0.015 mag, leading to a composite estimate of $E(B-V)$ = 0.015 ± 0.005 mag, which corresponds to an extinction of $A_V$ = 0.047 ± 0.016 mag. Alternatively, we can estimate the extinction by integrating a 3D extinction map[51] out to the distance of b Cen. This gives $A_G$ = 0.109 ± 0.014 mag, implying a colour slightly higher excess of $E(B-V)$ = 0.034 ± 0.005. Both values are very low relative to the photometric uncertainties in our analysis and do not impact the results. We adopt the former estimation, $A_V$ = 0.047 ± 0.016 mag, based on consistency in the literature. Such levels of extinction are normal for high-mass stars in the Scorpius-Centaurus region. While the properties of b Cen A can be relatively easily determined, b Cen B is invisible to most observing facilities, since it is both fainter than b Cen A and located close to it. The binarity is observed primarily based on the dynamical impact of b Cen B on b Cen A, through radial velocity variability[8] and excess astrometric motion[10], but there is insufficient dynamical data to fit an orbit to the observed motion. In addition, there is an interferometric data point from 2010[52], which appears to resolve the system at a separation (projected along a single baseline) of 9 mas. However, since the observation is based on a single epoch with a single baseline, and the detection is close to the instrument performance limit (M. Ireland, priv. comm.), it should be considered as a possible rather than a definitive detection. For the purpose of isochronal mass determinations of the stellar system, we have therefore adopted two edge case scenarios that define the envelope of possible masses for the central b Cen pair. In one edge case, we consider the possibility that the flux and mass of b Cen B are small enough to be effectively negligible, performing the isochronal analysis of b Cen A as if it was a single star. In the other edge case, we adopt the interferometrically derived brightness of b Cen B in order to estimate its mass in conjunction with the mass determination of b Cen A.

In all cases, we use isochrones based on the 15 Myr age of the system, using the PARSEC models[9] in the R-band range, since that is the wavelength band on which the interferometric measurement was centred. The first edge case in which b Cen A is treated as a dominant star in the system leads to a mass of 6 solar masses, while the second edge case gives individual masses for b Cen A and B of 5.6 and 4.4 solar masses respectively; i.e., a total system mass of 10 solar masses. Given that b Cen B is most likely somewhere in between these extremes, we can therefore adopt a total system mass range of 6—10 solar masses. In the scenario where b Cen A is the fully dominant component, there is a discrepancy of approximately 0.2 solar masses between using the R-band magnitude versus using the effective temperature as input for the isochronal analysis. In other words: If treated as a dominant component, b Cen A is mildly overluminous for its effective temperature, relative to isochronal expectations. This could be naturally explained if b Cen B does indeed contribute to the total flux, instead of being fully sub-dominant. However, it may also reflect uncertainties in the isochronal analysis itself. For example, the b Cen A has a projected rotational velocity of 129 km/s[53], which is relatively rapid. This distorts its shape and temperature distribution, such that its observed properties can vary depending on which direction it is observed from. Since we do not have this information, this constitutes an intrinsic uncertainty reflected in the differing outcomes. In the future, dynamical mass measurements of the system would eliminate these uncertainties, yielding robust and model-independent masses.

**Astrometric analysis.** When assessing whether a directly imaged planet candidate is a real companion or a background contaminant, a fundamental step is to verify that they share a common proper motion with the host star system, and are kinematically distinct from the population of potential contaminants. Each pairing of our three epochs shows statistically significant evidence for common proper motion (as well as orbital motion). Since the two BEAST epochs from 2019 and 2021 are the most readily reproducible data points, we use them as a baseline criterion for testing common proper motion, and remark that the formal significance is higher still if the 2000 epoch data point is also considered. The hypothesis that b Cen (AB)b is a static background contaminant can be rejected at the 14.2σ level based on its motion from the 2019 to the 2021 epoch. This conclusion remains robust if we allow for the hypothesized background contaminant to have a proper motion of its own. The proper motion scatter in the background population toward Sco-Cen is approximately 7 mas/yr in each direction[54]. Meanwhile, b Cen exhibits a proper motion of 43.7 mas/yr in total. This is a higher proper motion than most Sco-Cen targets, in part due to the fact that b Cen is on the "near" side of Sco-Cen, i.e., relatively close to us as observers. A background contaminant with a sufficiently high proper motion to keep up with b Cen can therefore be rejected at the 6.0σ level, and a contaminant keeping up with b Cen (AB)b can be rejected by 7.3σ, since the instantaneous orbital motion of b Cen (AB)b takes it in almost the opposite direction from the static background solution. By contrast, the two other point sources observed in the field around b Cen are both well consistent with a static background hypothesis. We fit orbital parameters for the orbital motion across the 21 year baseline using the "orbits for the impatient" (OFTI) module[15] within the "orbitize" code[55], and check the results against an MCMC code[56] which gives consistent results.

The planetary astrometry is referenced with respect to the position of its parent star. If the central point source is an unresolved binary, it will exhibit astrometric jitter over time, due to the orbital motion of the stellar pair. If the jitter is large enough, it can impose substantial noise into the apparent astrometry of the planet, and affect the orbital fitting. However, such an astrometric binary motion would be detected by wide-angle astrometric missions such as Hipparcos[44] and Gaia[10]. In the case of b Cen, the expected jitter is small, due to the close orbit of the binary. Indeed, both the differences in astrometry between Hipparcos and Gaia, as well as the intrinsic scatter within each catalogue, are less than 1 mas. This is several times smaller than the astrometric error bars for b Cen (AB)b, and therefore not a dominant source of error.

**Photometric and isochronal analysis.** We derive photometric values in the four measured spectral bands (J2, J3, K1, K2) using the characterisation arm of SpeCal[40] high-contrast pipeline, using the unsaturated point spread function of the star for fitting and determining the star-planet contrast, and calibrating against the star's 2MASS[57] near-infrared brightness. The 2MASS photometry has relatively large uncertainties due to the brightness of the star (approximately 4.5 mag in the near-infrared range), but it is consistent with the expected photometry for a B3V-type star. In the future, a more precisely measured stellar near-infrared photometric set independent from 2MASS would be useful, since the star is used as photometric reference for the planet, and therefore the photometric precision for the planet is partly affected by the precision for the star. We use the photometric values to derive an estimate for the bolometric luminosity $L_{bol}$ for the planet. For this purpose, we compare each photometric band individually to isochronal models based on both COND[58] and DUSTY[59] theoretical spectra, for ages of both 10 and 20 Myr. The idea behind this procedure is to represent the composite uncertainty set by uncertainties in the underlying model, in the age, and in the choice of spectral band. The method also implicitly includes the uncertainty in distance, since this is accounted for in the conversion between measured apparent magnitude and the absolute magnitudes used in the isochronal analysis. We then take the mean of all $L_{bol}$ values corresponding to each best-fit mode to represent the best-matching luminosity, and the standard deviation to represent the uncertainty. This gives $\log(L_{bol}/L_{sun})$ = -3.98 ± 0.19. We use this value along with the 15 ± 2 Myr age of the system to derive a mass estimate of 10.9 ± 1.6 Jupiter masses based on a range of initial entropies of the planet, as in ref.[13]. The result is consistent with other models, including the newly developed set of model tracks "planetsynth"[60] which favours a slightly lower mass of 8.9 ± 0.5 Jupiter masses.

**Formation analysis.** In the future, with high-resolution spectroscopy, it may be possible to derive clues about the formation scenario of b Cen (AB)b from chemical signatures, such as the atmospheric C/O ratio[61]. In the meantime, the most concrete traces of its origins are attained from its physical and orbital properties, and their relations to the broader population of detected gas giant planets. While wide-orbit (larger than ~10 au) giant planets are rare in total, with a frequency at the few per cent level or lower[62], they are increasingly found in direct imaging surveys, not least in the Sco-Cen region[63,64,65]. More massive stars are more likely to host wide giant planets in imaging surveys[16,17]. The companion frequency appears to increase with decreasing mass for the directly imaged exoplanet population[66], implying a separate formation channel from stellar companions. Statistical investigations of the radial velocity exoplanet population have shown that companions in the mass range of ~4 Jupiter masses and lower show a correlation with metal-enrichment in their parent stars while companions in the range of ~10 Jupiter masses and higher do not[67], which in that context seems to imply that companions in the same mass range as b Cen (AB)b preferentially form through a star-like channel. However, these metallicity-mass relations are derived for Sun-like stars, whereas B-stars are several times more massive, and with consequently more massive disks, may be expected to host multiple times more massive planets on average as well. A few of the imaged planets are circumbinary[35,68], and there is so far no statistically significant difference between single and binary stars in terms of the probability to host wide giant planets[26].

The mass ratio to the central stellar pair of 0.0011 and projected separation of 560 au distinguishes b Cen (AB)b within the exoplanet population, as shown in Fig. 2. In particular, the mass ratio is similar to that of the 51 Eri system[69], and significantly smaller than other directly imaged planetary systems. Such a low mass ratio is indicative of a planetary formation scenario, distinct from a scenario in which b Cen (AB)b would have represented the low-mass end of a stellar population, as a tertiary component in the b Cen system[70]. In order to quantify this, we have integrated a model that includes both populations and compared the amplitude of each population. The model spans mass ratios between 0.0005 and 0.02, and orbital separation ranges between 250 and 1000 au. The brown dwarf companion model is based on an extrapolation of a companion survey around intermediate mass stars[71]. We adopt their log-normal orbital distribution with a peak near 400 AU, the companion mass ratio distribution of ref.[14], and normalize with the observed frequencies between 30—800 au from $0.1 < q < 1.0$[71]. For the planet population, we fitted a log-normal distribution to published frequencies of gas giant planets 5—13 Jupiter masses ($0.0025 < q < 0.0065$) from 0—320 AU (with a peak between 1—10 au) based on radial velocity and direct imaging surveys[16,72,73,74,75]. The model explicitly accounts for variations in companion frequency as a function of mass ratio and orbital separation. A similar model was used to interpret data from the SHINE survey[17] and is described in detail in Meyer et al. (submitted). The resulting probabilities are $3.1 \times 10^{-5}$ for the stellar scenario and $3.8 \times 10^{-3}$ for the planetary scenario; i.e., the probability for a planetary origin is more than 100 times larger than a stellar origin. Both probabilities are low, which reflects the general scarcity of very low-mass wide companions to high-mass stars. We emphasize that these models are necessarily based on extrapolations, since the high-contrast circumstellar environment of B-type stars has not been statistically explored in detail yet, so they should be regarded as tentative. Indeed, the most relevant survey for assessing population distributions at low masses around B-type stars will be the BEAST survey itself. When the survey is finished, the prospects for robustly evaluating how b Cen (AB)b fits into the larger exoplanet demographics will therefore greatly improve.

An aspect of the possible disk instability formation path that remains unclear is to which extent it can reproduce the mass of b Cen (AB)b. Fragments produced in theoretical work are typically more massive than 10 Jupiter masses for disks around early-type stars[76,77,78]. Possible pathways to planetary masses may include formation at a relatively late stage when the disk is partially dispersed, or as previously discussed, formation elsewhere in the disk followed by a migration and scattering process, possibly involving tidal downsizing[79,80] for reducing the embryo mass.

**Code availability.** Data were processed using recipes at the SPHERE Data Center. Access to the Data Center is available by following the instructions at https://sphere.osug.fr/spip.php?article47&lang=en.

**Data availability.** All the raw data used in this study are available at the European Southern Observatories archive (http://archive.eso.org/cms.html) under program ID 1101.C-0258, by default after a proprietary time of one year after each respective data set was acquired, but earlier access can be provided upon reasonable request to the corresponding author. Processed data are available from the Data Center by following the instructions at https://sphere.osug.fr/spip.php?article74&lang=en.

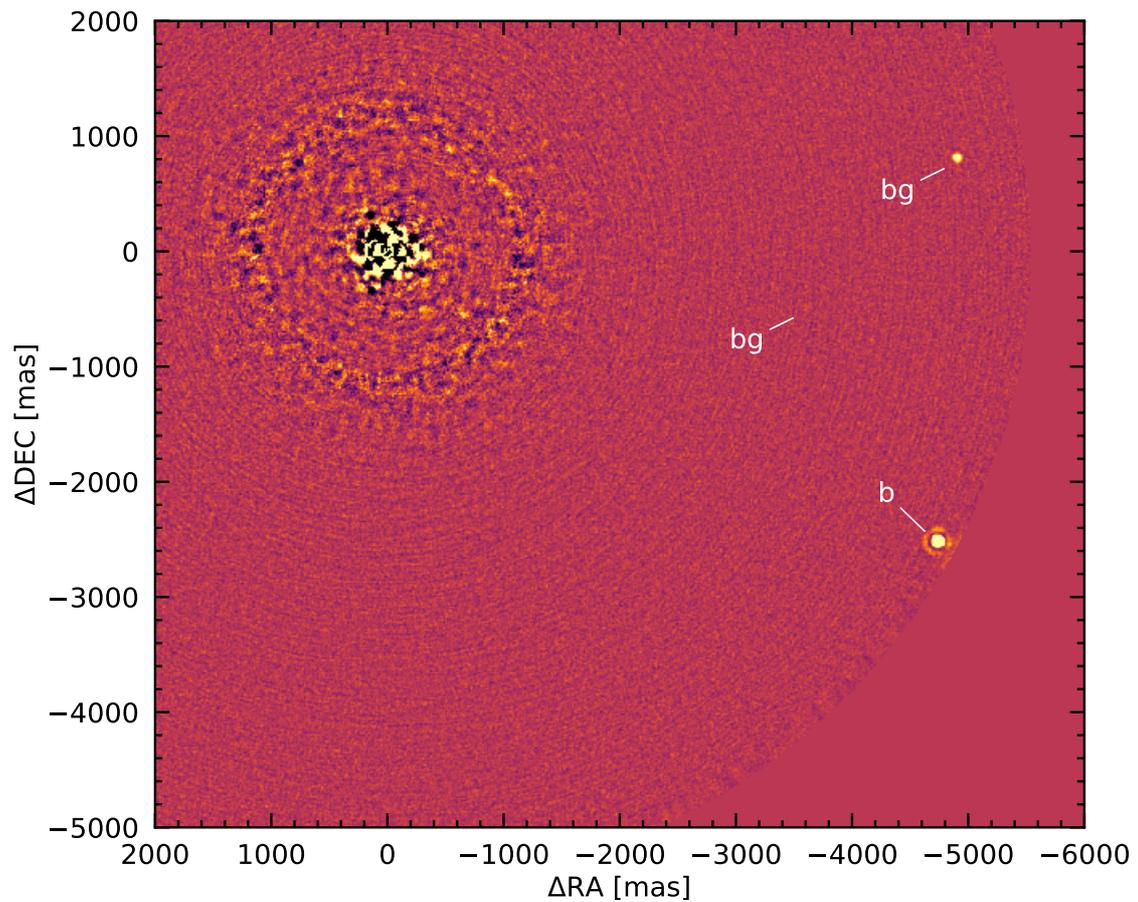

**Figure 1 | Image of b Cen (AB)b.** The planet itself is denoted 'b' and located near the edge of the image. The central pattern is residual noise from the light of the parent stellar system, which has been blocked by a coronagraph as well as digitally subtracted, in this case using so-called classical angular differential imaging. The two background stars also visible in the field are both denoted 'bg'. The image is in the K1-band. North points up and East points to the left in the image.

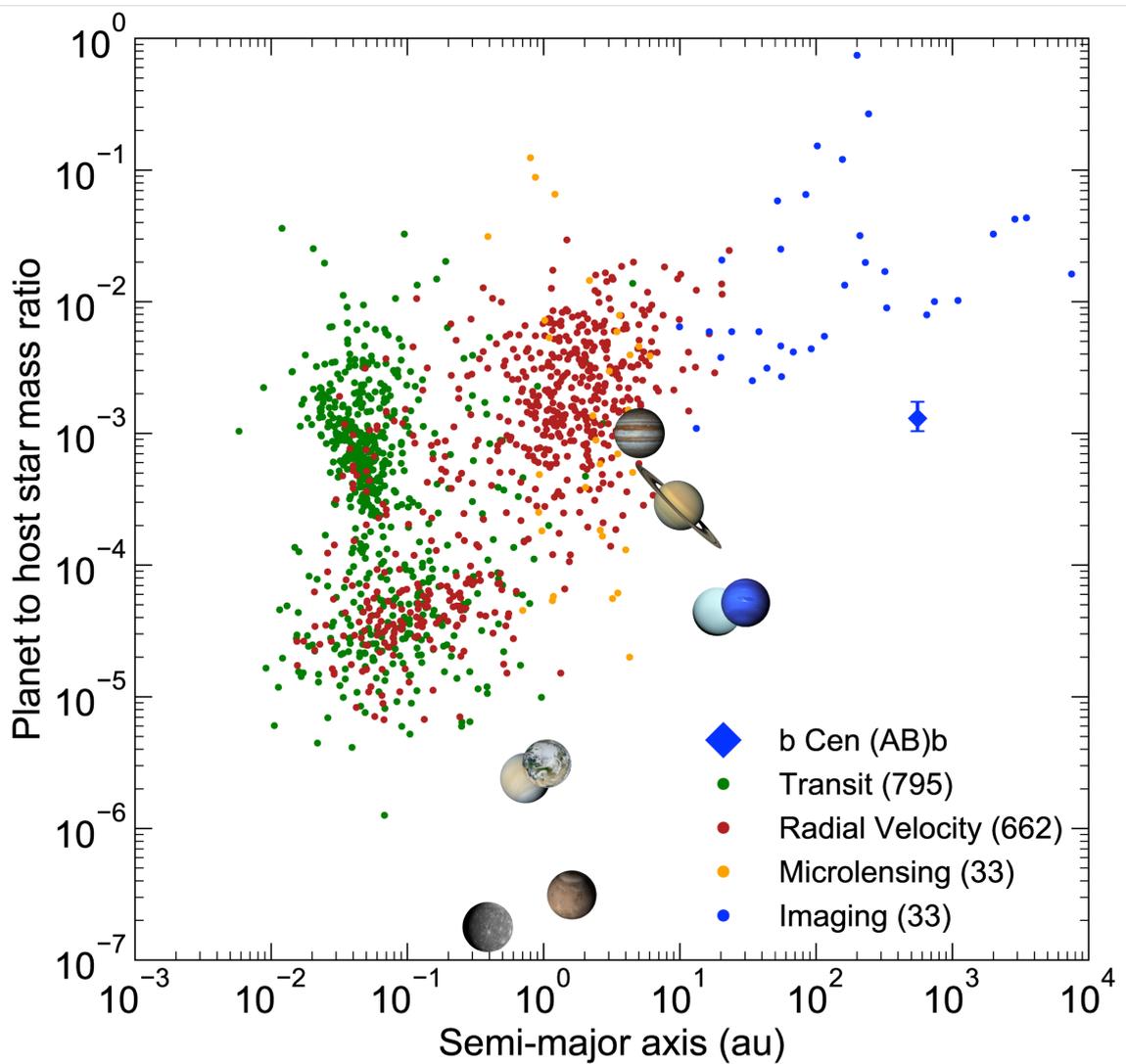

**Figure 2 | The planet-to-star mass ratio of b Cen (AB)b in an exoplanetary context.** Small circles are confirmed exoplanets with parent stellar masses known to better than 30% precision, retrieved from the NASA Exoplanet Archive. Both single and binary star systems are included. The planets are colour coded by detection method, where green circles are transit detections, red are radial velocity detections, black are microlensing detections, and blue are imaging detections. The Solar system planets (images from NASA) are also plotted for reference. The diamond symbol denotes b Cen (AB)b, which has an unusually low mass ratio to the central system relative to other detected planets in the wide, directly imaged population.

**Table 1 | Parameters for the b Cen system**

|  | Value | SI units |
|---|---|---|
| **Physical parameters** | | |
| Star A mass | 5—6 $M_{Sun}$ | 1.0—1.2×10$^{31}$ kg |
| System mass | 6—10.0 $M_{Sun}$ | 1.2—2.0×10$^{31}$ kg |
| Planet mass | 10.9 ± 1.6 $M_{Jup}$ | 2.0×10$^{28}$ kg |
| System age | 15 ± 2 Myr | 4.7×10$^{14}$ s |
| Distance | 99.7 ± 3.1 pc | 3.1×10$^{18}$ m |
| **Orbital parameters** | | |
| Projected separation (2019) | 556 ± 17 au | 8.3×10$^{13}$ m |
| Eccentricity | <0.40 | |
| Inclination | 128°—157° | |
| Orbital period | 2650—7170 yr | 0.8—2.3×10$^{11}$ s |

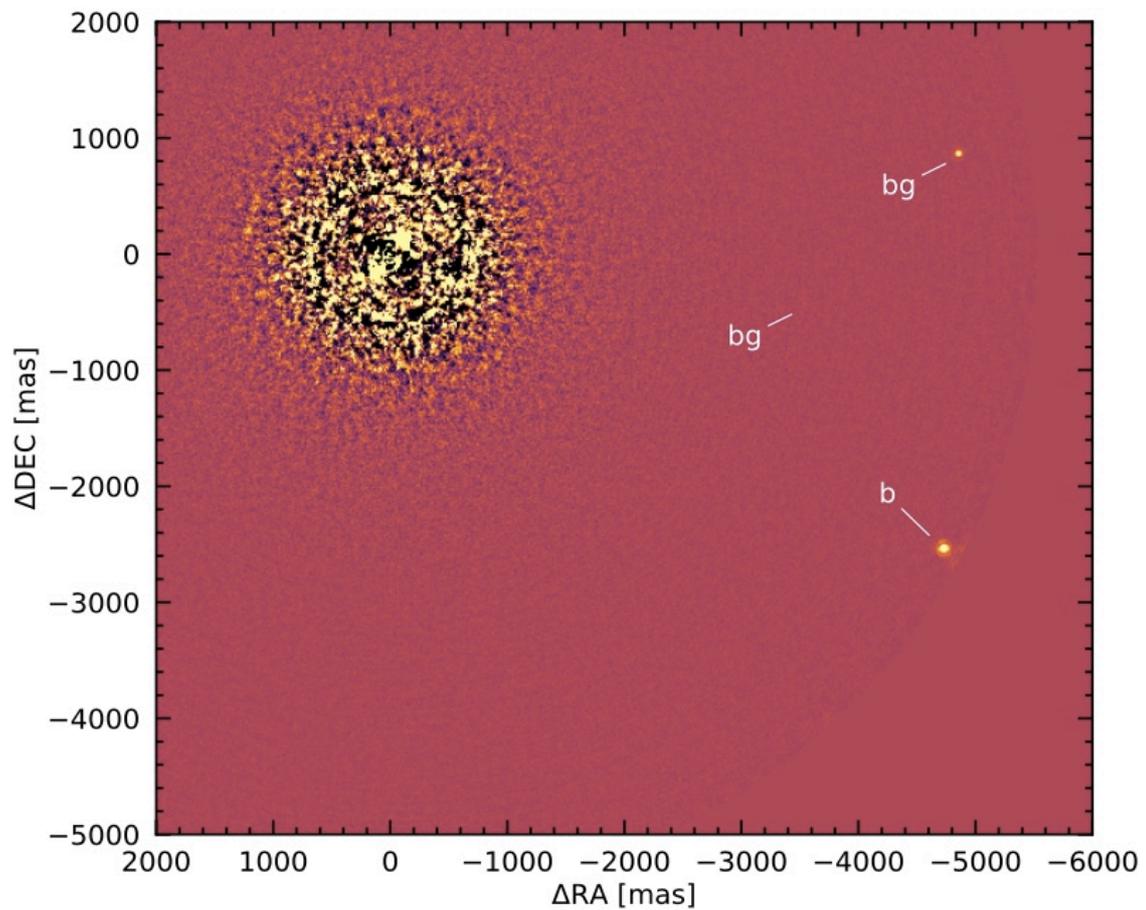

**Extended data figure 1 | J2-band image of b Cen (AB)b.** The image reduction is performed with classical angular differential imaging. The planet is denoted 'b' and the brighter of the background stars is denoted 'bg'. The fainter background star cannot be easily seen at the contrast/saturation of this display, which is chosen to optimize visibility of other image elements.

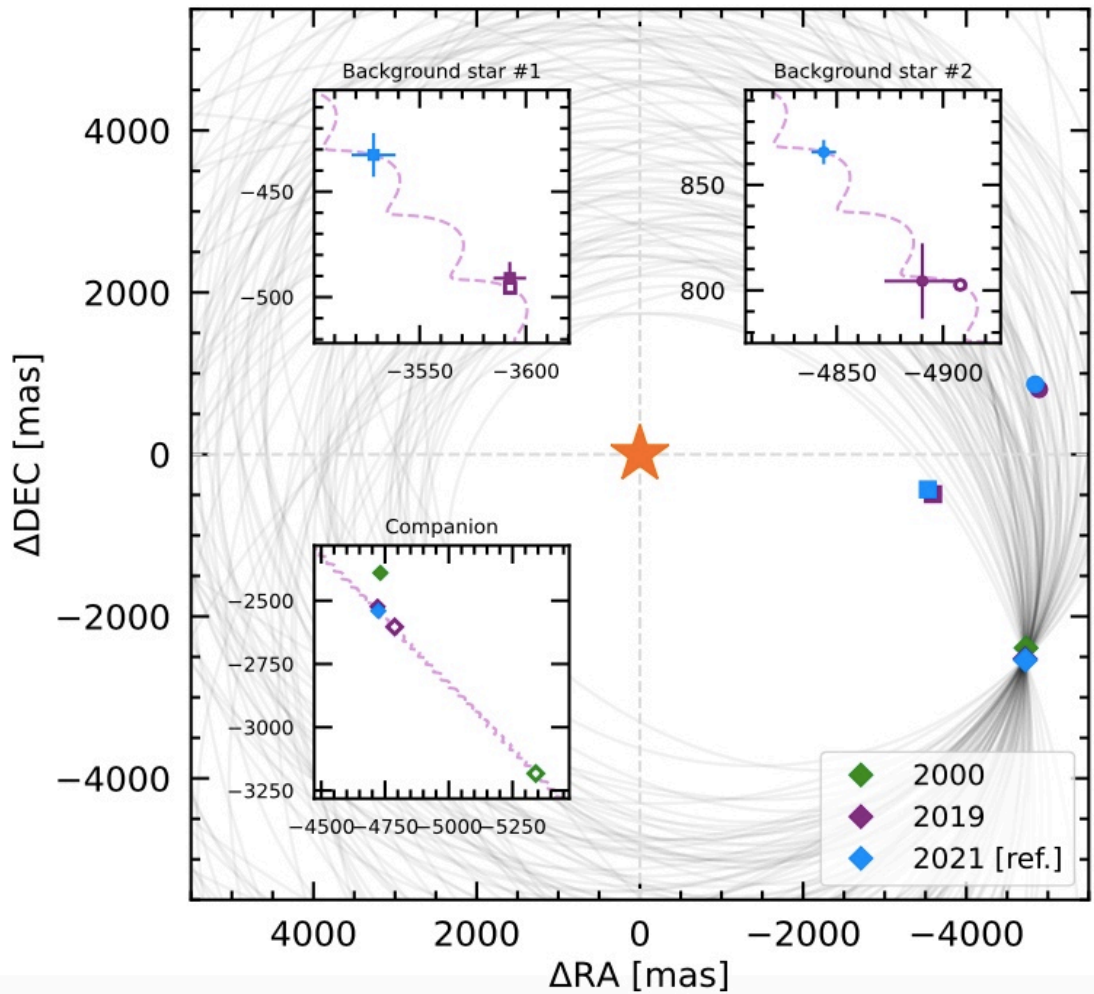

**Extended Data Figure 2 | Astrometric motion of b Cen (AB)b and the background stars.** The image shows the astrometric motion of the three point sources detected around b Cen, in the reference frame of b Cen itself. Squares show the locations of background star 1 at epochs 2019 (purple) and 2021 (blue). Circles show the locations of background star 2 at the same epochs. Diamonds show the locations of b Cen (AB)b, both at the 2019 and 2021 epochs, but also in the 2000 epoch (green) where it could additionally be retrieved. Gray tracks show a representative collection of orbits that fit the observed motion of b Cen (AB)b. The insets zoom in on the locations around background star 1 (upper left inset), background star 2 (upper right inset), and the confirmed planet b Cen (AB)b (lower left inset). The filled symbols are the measured locations, while the open symbols show the projected motion expected for a static background object (which would follow the dashed trajectories over time), where 2021 is chosen as the reference epoch.

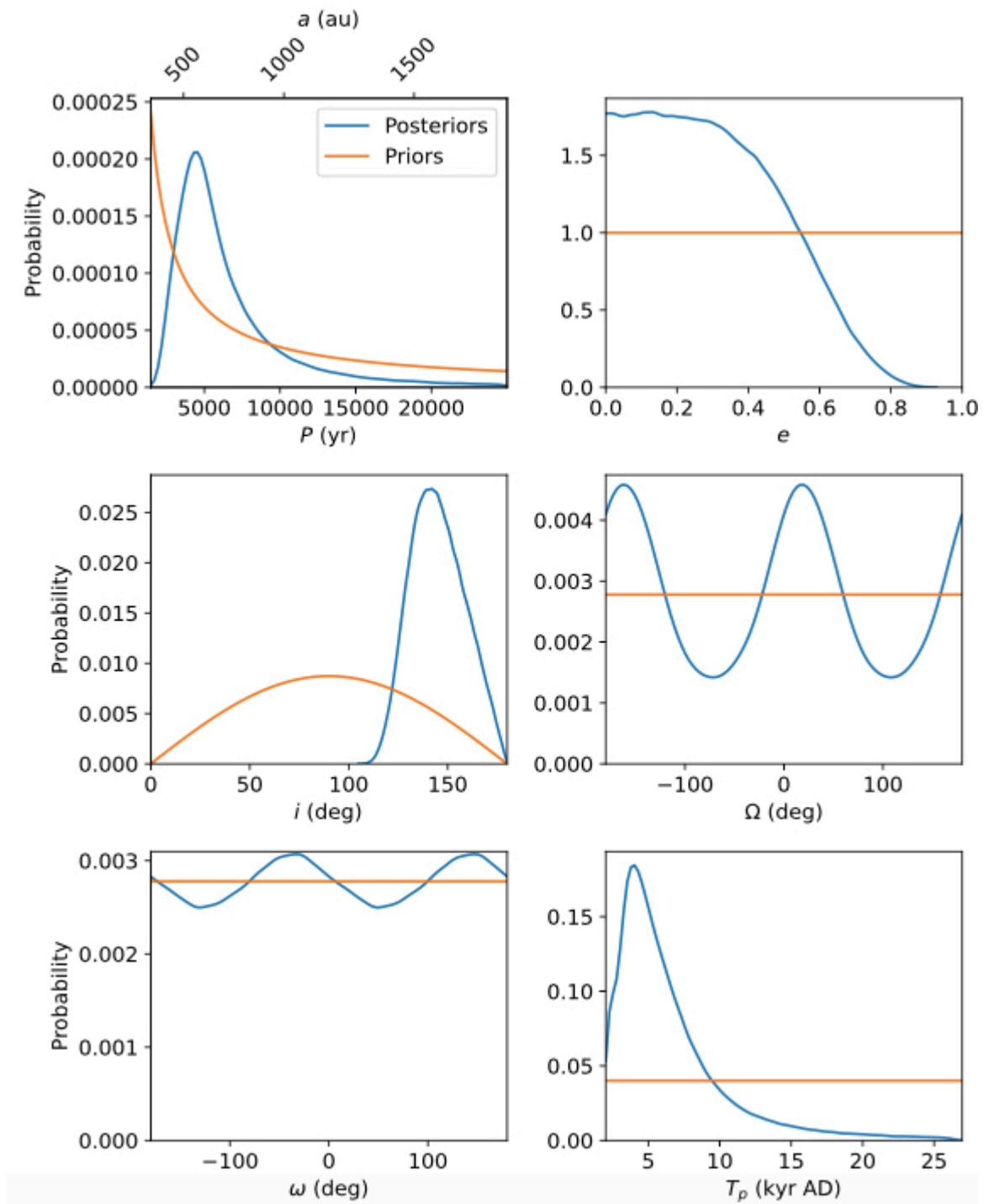

**Extended Data Figure 3 | Orbital parameters of b Cen (AB)b.** Prior (in orange) and posterior (in blue) distributions for the full set of orbital parameters: Orbital period $P$, eccentricity $e$, inclination $i$, ascending node $\Omega$, argument of periapsis $\omega$, and time of periapsis $T_p$.

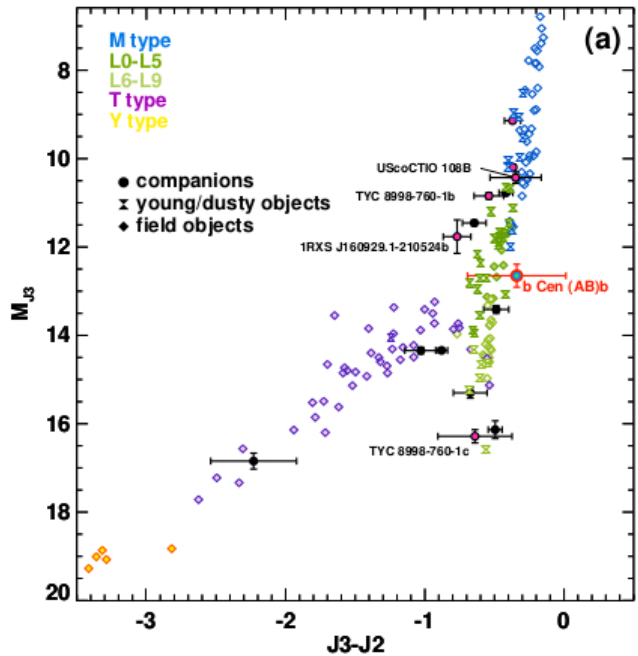

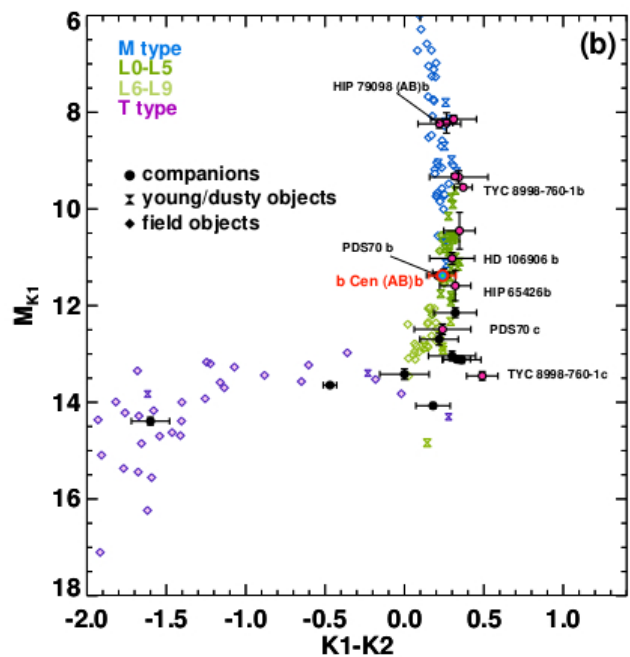

**Extended Data Figure 4 | Colour-magnitude diagrams for b Cen (AB)b. a,** J2-J3 colour versus absolute J2 magnitude. **b,** K1-K2 colour versus absolute K1 magnitude. The planet b Cen (AB)b is plotted as a blue-green star, and follows the same colour trends as are generally observed for young planetary and substellar companions to stars, plotted as purple and black symbols with error bars. Symbols without error bars are young and field brown dwarfs.

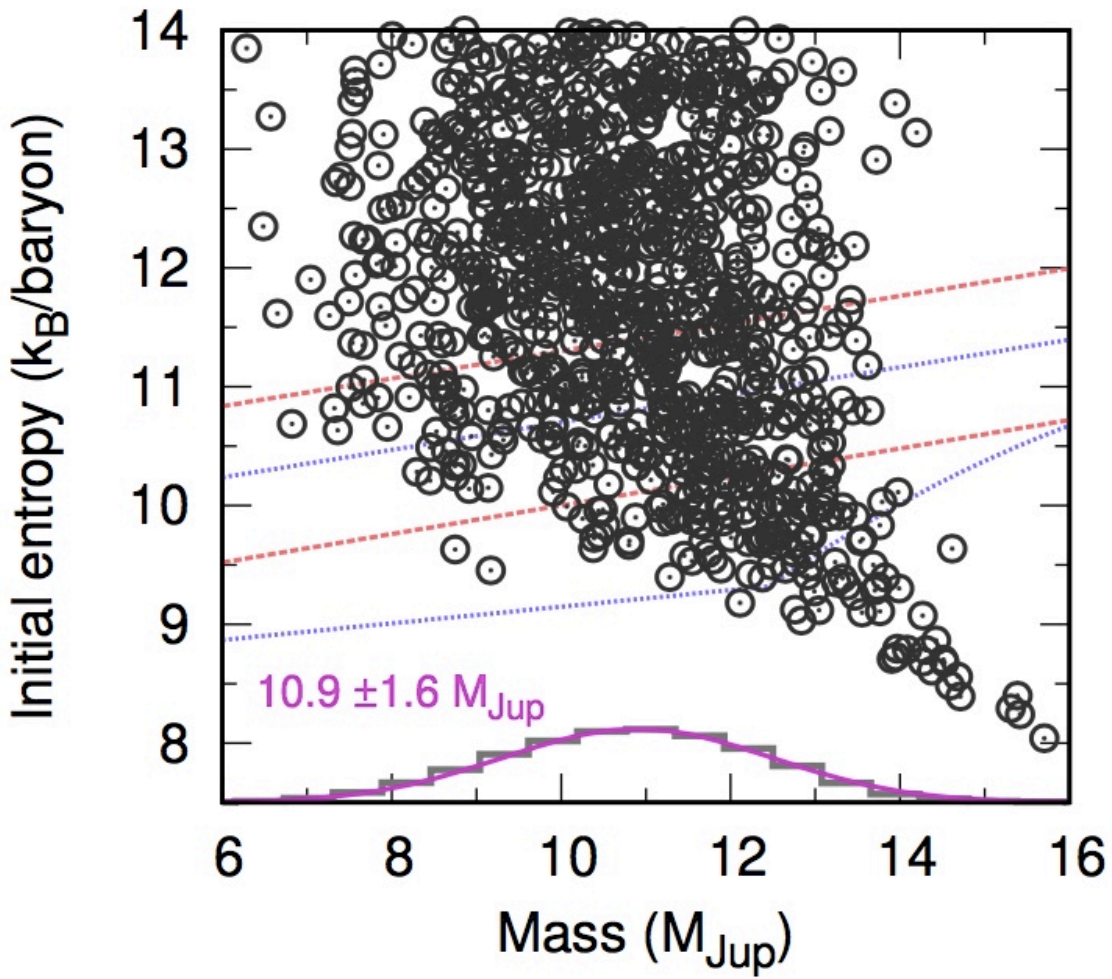

**Extended Data Figure 5 | Constraints on mass and initial entropy for b Cen (AB)b.** Posterior probability distribution for the mass and initial entropy of b Cen (AB)b based on its brightness and age. The BEX-Cond models are used in this MCMC exploration, but the results are not very sensitive to the choice of the atmospheric model since the bolometric luminosity (and not a magnitude) is used. The red dotted (blue dashed) lines show for reference the approximate minimum and maximum of the hot-start (cold-start) planets in the Bern population synthesis[81]. A subset of the models is shown for plotting purposes.

|  | Separation (mas) | Position angle (°) |
| --- | --- | --- |
| **b Cen (AB)b** | | |
| 51690 MJD | 5301 ± 53 | 243.1 ± 0.1 |
| 58563.33 MJD | 5351.76 ± 3.66 | 241.872 ± 0.034 |
| 59315.27 MJD | 5362.04 ± 6.69 | 241.722 ± 0.068 |
| **Faint background star** | | |
| 58563.33 MJD | 3625.80 ± 14.99 | 262.217 ± 0.152 |
| 59315.27 MJD | 3552.06 ± 8.68 | 263.064 ± 0.134 |
| **Bright background star** | | |
| 58563.33 MJD | 4954.49 ± 3.74 | 279.346 ± 0.035 |
| 59315.27 MJD | 4919.19 ± 6.08 | 280.124 ± 0.067 |

**Extended Data Table 1 | Astrometric values for point sources around b Cen.**

|  | Apparent (mag) | Absolute (mag) |
|---|---|---|
| **b Cen (AB)b** | | |
| J2 (1190 nm) | 17.98 ± 0.25 | 12.99 ± 0.26 |
| J3 (1273 nm) | 17.64 ± 0.25 | 12.65 ± 0.26 |
| K1 (2110 nm) | 16.37 ± 0.06 | 11.38 ± 0.08 |
| K2 (2251 nm) | 16.13 ± 0.06 | 11.14 ± 0.08 |
| **Faint background star** | | |
| J2 (1190 nm) | 23.02 ± 0.48 | |
| J3 (1273 nm) | 23.14 ± 0.52 | |
| K1 (2110 nm) | 20.90 ± 0.18 | |
| K2 (2251 nm) | Not detected | |
| **Bright background star** | | |
| J2 (1190 nm) | 19.90 ± 0.25 | |
| J3 (1273 nm) | 19.62 ± 0.25 | |
| K1 (2110 nm) | 18.50 ± 0.06 | |
| K2 (2251 nm) | 18.40 ± 0.08 | |

**Extended Data Table 2 | Photometric values for point sources around b Cen**